\documentclass[12pt]{article}

\usepackage{amsmath}
\usepackage{hyperref}
\usepackage{mathtools}
\usepackage{amsfonts}
\usepackage{graphicx}
\usepackage[numbers]{natbib}
\usepackage{tikz}
\usepackage{amsthm}
\usepackage{verbatim}

\newcommand{\RR}{\mathbb{R}}

\newcommand{\KK}{\mathcal{K}}
\newcommand{\ef}{E_e}
\newcommand{\ec}{E_c}
\newcommand{\mr}{\mathring}
\newcommand{\YY}{\mathcal{Y}}
\newcommand{\kk}{\kappa}
\newcommand{\lich}{\mathrm{Lich}}
\newcommand{\bigo}{\mathcal{O}}
\newcommand{\WW}{\mathcal{W}}
\newcommand{\QQ}{\mathcal{Q}}
\newcommand{\sss}{\mathcal{S}_\sigma}
\newcommand{\LL}{\mathcal{L}}
\newcommand{\vep}{\varepsilon}
\newcommand{\tr}{\mathrm{tr}}
\newcommand{\di}{\mathrm{div}}
\newcommand{\cvl}{-\Delta_{\mathbb L}}

\newcommand{\mrcvl}{\mathring{\mathrm{div} L}}
\newcommand{\coker}{\mathrm{coker}}
\newcommand{\ran}{\mathrm{ran}}
\newcommand{\vt}{\vartheta}
\newcommand{\mcs}{\mathcal{S}}

\newtheorem{thm}{Theorem}[section]
\newtheorem{lem}[thm]{Lemma}
\newtheorem{prop}[thm]{Proposition}
\newtheorem{cor}[thm]{Corollary}

\title{Non-constant mean curvature trumpet solutions for the Einstein constraint equations}
\author{Jeremy Leach}
\date{}

\begin{document}

\maketitle

\begin{abstract}
We prove the existence of a large class of initial data for the vacuum Einstein equations which possess a finite number of asymptotically Euclidean and asymptotically conformally cylindrical or periodic ends. Aside from being asymptotically constant, only mild conditions on the mean curvature of these initial data sets are imposed. 
\end{abstract}


\section{Introduction}

There has been a great deal of recent progress in the study of the initial value problem in general relativity, and much of this progress has come from the conformal method. This technique, described in the following section, has proved to be an extremely useful way to parametrize the set of solutions to the Einstein constraint equations on closed manifolds and manifolds possessing asymptotically Euclidean or hyperbolic ends. In fact, it was shown quite recently by Maxwell \cite{max1} that, at least in the case of closed manifolds, the set of solutions for the constraint equations one obtains from the conformal method is identical to the set parametrized by several well known competing parametrization schemes (such as the conformal thin-sandwich method).

Some of the most interesting known solutions to the Einstein field equations, however, possess spacelike slicings by manifolds which have completely different asymptotic behavior. Perhaps the most notable example is the extreme Kerr black hole, which is a stationary rotating black hole whose total angular momentum $|a|$ is, with appropriately chosen units, equal to its total mass $m$. This solution is known to admit a maximal time slicing by manifolds diffeomorphic to $\RR^3 \setminus 0$ which are asymptotically Euclidean as $r\to\infty$ and asymptotically conformally cylindrical as $r\to 0$. While such black holes have never been detected in nature, there is empirical evidence for the existence of non-extreme black holes throughout the universe with $|a|/m >0.98$ \cite{astro}. On the other hand, one does not expect to find ``hyper-extreme'' ($|a|/m > 1$) Kerr black holes in nature because these would possess naked singularities in violation of the weak cosmic censorship conjecture. One expects a universe possessing naked singularities to fail to be globally hyperbolic, possibly admitting closed timelike curves. In this sense, extreme Kerr black holes lie at the threshold of what is known to be physical and what is theorized to be unphysical. For this reason, the extreme Kerr solution is of great interest to physicists and mathematicians alike.

In general, we will say that a Riemannian manifold with a finite number of asymptotically Euclidean and asymptotically conformally cylindrical or periodic ends is of \textit{trumpet type}, and we will make these definitions precise in the next section. There have been a number of recent efforts to construct initial data sets with such asymptotics. For example, a large family of Bowen-York initial data was constructed in \cite{bow}, and small perturbations of extreme  Kerr initial data were produced in \cite{dg}. The aim of this paper is to construct, via the conformal method, a large class of initial data sets with trumpet geometry. As with most existence results to date, some restrictions are imposed on the mean curvature of these initial data sets. To be precise, let $(M,g)$ be a Riemannian $n$-manifold ($n \geq 3$) and let $K$ be a symmetric 2-tensor on $M$. We call a triple $(M,g,K)$ an \textit{initial data set} for the vacuum Einstein field equations with cosmological constant $\Lambda$ if the following equations are satisfied:
\begin{align}\label{consteq}
R_g - |K|^2 _g + (\tr_g K)^2 &= 2\Lambda \\
\di_g K - \nabla(\tr_g K) &= 0.
\end{align}

One can often find a solution to these equations by first fixing a non-physical background metric $g$ on $M$, and then looking for a solution to the constraints $(\tilde g, \tilde K)$ which has the form
\begin{align}
\tilde g &= \phi^\kk g \label{conftrans1}\\
\tilde K &= \frac{\tau}{n} \tilde g + \phi^{-2}(\sigma + LW)\label{conftrans2}
\end{align}

\noindent where $\phi$ and $\tau$ are scalar functions, $\sigma$ is a 2-tensor which is traceless divergence-free with respect to $g$, $W$ is a vector field and $\kk$ is a dimensional constant given by $\kk = 4/(n-2)$. The operator $L$ is the conformal Killing operator, defined as
\[
(LX)_{ij} = \nabla_i X_j + \nabla _j X_i - \frac2{n} (\di_g X) g_{ij}.
\]

\noindent Here $\nabla$ is the Levi-Civita connection associated to the background metric $g$, and we shall refer to any nontrivial vector field $W$ which satisfies $LW =0$ identically as a \textit{conformal Killing field}. Note that the function $\tau$ becomes the trace of the physical extrinsic curvature tensor $\tilde K$, and so we refer to $\tau$ as the \textit{mean curvature} of the initial data set we construct. This construction of solutions $(\tilde g, \tilde K)$ to the constraints by specifying a triple $(g, \tau, \sigma)$ and solving for $(\phi, W)$ is known as the \textit{conformal method}. There are many detailed expositions on the conformal method  (see \cite{ise1}, for example), and the key observation of this approach is that the constraint equations are reduced to the coupled semilinear elliptic system
\begin{align}
\Delta_g \phi - c_n R_g \phi &= \beta \phi^{\kk-1} - c_n |\sigma + LW|_g ^2 \phi^{-\kk-1} \label{lcby1} \\
\di_g LW &= b_n \phi^\kk \nabla\tau \label{lcby2}.
\end{align}

\noindent Here $c_n$ and $b_n$ are dimensional constants given by $(n-2)/4(n-1)$ and $(n-1)/n$ respectively, and $\beta$ is a scalar function given by
\[
\beta = \frac{n-2}{4n} \tau^2 - \frac{n-2}{2(n-1)} \Lambda.
\]
When constructing initial data which have asymptotically Euclidean ends, as we do in this paper, one must assume $\Lambda = 0$. The first equation is known as the \textit{Lichnerowicz equation} and, as previous authors have, we shall refer to equations \eqref{lcby1}-\eqref{lcby2} together as the LCBY equations, in honor of Lichnerowicz, Choquet-Bruhat and York who were among the first to study them. We shall refer to the operator $\cvl := \di_g L$ as the \textit{conformal vector Laplacian} associated to $g$.\footnote{The minus sign in our notation is meant to conform with the convention of other authors that $\Delta_\mathbb{L} = L^* L = -\di_g L$. This sign was omitted in \cite{lea2}.}

The utility of the conformal method is that it will allow us to construct solutions to the constraint equations $(\tilde g, \tilde K)$ for which the physical metric $\tilde g$ is of trumpet type and the physical extrinsic curvature $\tilde K$ has commensurate asymptotics on the ends. To do this, one simply specifies a background metric $g$ of trumpet type and a pair $(\tau, \sigma)$ with reasonable asymptotic assumptions and seeks a solution $(\tilde \phi, \tilde W )$ chosen so that the resulting physical data $(\tilde g, \tilde K)$, as prescribed by \eqref{conftrans1}-\eqref{conftrans2}, remain of trumpet type. Defining the triple $(g,\tau, \sigma)$ to be of \textit{trumpet type} if $\tau$ and $\sigma$ satisfy these asymptotic assumptions (stated precisely in the next section), we may state our main existence theorem:

\begin{thm}\label{mainthm} Let $(g,\tau,\sigma)$ be Yamabe positive conformal data of trumpet type such that $g$ admits no conformal Killing fields, and suppose that $\sup_M |\sigma|_g$ is small compared to $\|\nabla\tau\|^{1-n} _{0,-\delta}$. The system \eqref{lcby1}-\eqref{lcby2} then has a solution $(\phi, W) \in C^{2,\alpha} \times C^{2,\alpha}$ such that the resulting physical metric has the same asymptotic type as the background metric $g$.
\end{thm}

\noindent In this theorem, the ``same asymptotic type'' means that if our background metric is asymptotically Euclidean, then the conformal factor $\phi$ can be chosen so that the physical metric $\tilde g$ is asymptotically Euclidean as well, and similarly for ends of cylindrical type. Here $\|\cdot\|_{0,-\delta}$ is a weighted sup-norm and $C^{2,\alpha}$ are H\"older spaces (both defined below). 

Whereas early existence theorems for the LCBY equations assumed a constant mean curvature (CMC) $\tau$, this theorem is an analogue of a number of recent non-CMC existence results such as those for initial data sets with a finite number of asymptotically conformally cylindrical ends or asymptotically periodic ends \cite{lea2}, a finite number of asymptotically Euclidean ends \cite{dimm}, and for those defined on closed manifolds \cite{hnt}, \cite{max2}. Namely, a smallness condition on the tensor $\sigma$ is imposed, like the one given in Theorem \ref{mainthm}, in place of a CMC or near-CMC condition. Note, however, that the conditions of this theorem appear to be compatible with near-CMC existence results. This is because a typical near-CMC condition imposes a smallness condition on $|\nabla\tau|$. In this case one would expect the quantity $\|\nabla\tau\|^{1-n} _{0,-\delta}$ to be quite large. Hence an arbitrary choice of $\sigma$ would satisfy the hypothesis of Theorem \ref{mainthm} given a sufficiently stringent near-CMC condition. This leads the author to suspect that the set of solutions to the constraints one obtains from near-CMC conditions is the same as those obtained from a ``small $\sigma$'' assumption. 


\section{Notation and definitions}

For simplicity, we will henceforth only consider manifolds with exactly one asymptotically Euclidean end and one end of cylindrical type. More precisely, consider an $n$-manifold $M$ which possesses a compact subset $\KK$ with smooth boundary such that
\begin{equation*}
M\setminus \KK = \ef \sqcup \ec.
\end{equation*}

\noindent Here $\ef$ is diffeomorphic to $\RR^+ \times S^{n-1}$ and $\ec$ is diffeomorphic to $\RR^+ \times \Sigma$ for some closed $(n-1)$-manifold $\Sigma$. Suppose $g_0$ is a metric on $M$ which is exactly Euclidean on $\ef$ and exactly conformally cylindrical on $\ec$. This is to say that there are coordinate functions $r$ and $x$ such that $r$ is identically 2 on $\partial \ef$ and
\[
g_0|_{\ef} = dr^2 + r^2 g_{S^{n-1}}.
\]

\noindent Similarly $x$ is identically 2 on $\partial \ec$ and
\[
g_0|_{\ec} = \mr\psi^{\kk-2}(dx^2 + g_\Sigma)
\]

\noindent where $\mr\psi$ is a positive function on $\Sigma$ and $g_\Sigma$ is some fixed metric on $\Sigma$. We use this model metric to define doubly weighted function spaces as follows. We first define, for any $1 \leq p < \infty$, the space $L^p _{\mu,\nu}$ to be the space of all functions $u$ satisfying
\begin{equation*}
\|u\|_{p,\mu,\nu} := \left ( \int_M e^{-p \mu x}r^{-p\nu - n}|u|^p dV_{g_0} \right ) ^{\frac{1}{p}} < \infty.
\end{equation*}

\noindent Note in particular that $L^p = L^p _{0,-n/p}$. We then define the corresponding Sobolev spaces $W^{k,p} _{\mu,\nu}$ to be those functions $u$ satisfying
\begin{equation*}
\|u\|_{k,p,\mu,\nu} := \sum_{|\alpha| \leq k} \|r^{|\alpha|}D^\alpha u\|_{p,\mu,\nu} < \infty.
\end{equation*}

\noindent Here $D$ is the connection associated to $g_0$, and the corresponding Sobolev spaces of tensor fields are defined analogously. Now given a Riemannian metric $g$, we say that $g$ is of trumpet class $W^{k,p} _{\mu,\nu}$ if
\begin{equation*}
g - g_0 \in W^{k,p} _{\mu,\nu}.
\end{equation*}

\noindent We say that the metric $g$ is \textit{asymptotically conformally cylindrical} (ACC) on $\ec$ if $\mu<0$, and we say $g$ is \textit{asymptotically Euclidean} (AE) on $\ef$ if $\nu<-n$.  As usual, we denote by $H^k _{\mu,\nu}$ the space $W^{k,2} _{\mu,\nu}$. \noindent We also define the local H\"older norms, denoted $\|\cdot\|_{k,\alpha; B_1(q)}$, as
\[
\|X\|_{k,\alpha; B_1 (q)} = \sum^k _{j=0} \sup_{B_1(q)} |D^j X| + \sup_{p_1, p_2 \in B_1 (q)} \frac{|D^k X(p_1) - D^k X (p_2)|}{\mathrm{dist}_{g_0} (p_1,p_2)^\alpha}
\] 

\noindent where we have, without loss of generality, assumed that the injectivity radius of $(M,g_0)$ is one. Similarly, the global $C^{k,\alpha}$-norm is given by
\[
\|X\|_{k,\alpha} = \sup_{q\in M} \|X\|_{k,\alpha; B_1 (q)}.
\]

\noindent Of course, we can similarly define the weighted H\"older spaces on $M$ by setting $C^{k,\alpha} _{\mu,\nu}$ to be the space of all functions (or tensor fields) $u$ which satisfy
\begin{equation*}
\|r^{-\nu}e^{-\mu x} u\|_{C^{k,\alpha}}  < \infty,
\end{equation*}

\noindent and we say that $g$ is of trumpet class $C^{k,\alpha} _{\mu,\nu}$ if
\[
g - g_0 \in C^{k,\alpha} _{\mu,\nu}.
\]

\noindent We will also use the notation $\|\cdot\|_0$ to denote the usual sup norm.

Next we state that conformal data $(g,\sigma,\tau)$ is of \textit{trumpet type} if $g$ is of trumpet class $C^{3,\alpha} _{-1,-\gamma}$ for some constant $0 < \gamma < 2(n-2)$, and if, for fixed cutoff functions $\chi_e$ and $\chi_c$ which localize to $\ef$ and $\ec$ respectively, the quantities $\chi_e \tau, \chi_e \sigma, \chi_c \tau - \mr\tau$, and $\chi_c \sigma - \mr\sigma$ all belong to $C^{1,\alpha} _{-\delta,-\gamma/2-1}$, where $\delta$ is chosen as in the regularity theorem for the conformal vector Laplacian which is given below. Here $\mr\tau$ is a fixed nonzero constant, and $\mr\sigma$ is a fixed 2-tensor in $C^{1,\alpha}(\Sigma; S^2 _0(\Sigma)$. Note that we will sometimes refer to only the metric $g$ as being of trumpet type, though the context will make clear whether we are referring to $g$ or the full triple $(g,\tau,\sigma)$.

We shall also refer to conformal data $(g,\tau,\sigma)$ as being ``of trumpet type'' when the model metric $g_0$ is periodic on the end $\ec$. This is to say that $g_0|_{\ec}$ is the restriction to the half-cylinder of the lift of some Riemannian metric on the manifold $S^1 \times \Sigma$. We will denote by $T$ the $x$-period of the model periodic metric on $\ec$. In this case, $\mr\sigma$ is a tensor on $S^1\times \Sigma$. The analysis below is analogous for the two cases, and we will point out relevant differences below.

\begin{thm}\label{trumpetcvlmap}

Let $(M,g)$ be a Riemannian manifold of trumpet type which does not admit any bounded conformal Killing fields. Then there exists a $\delta_* >0$ such that, for any $0 < \delta < \delta_*$ and any $0 < \nu < n-2$, the map 
\begin{equation}\label{cvlsurj1}
\cvl : H^{k+2} _{\delta,-\nu} (M;TM) \rightarrow H^k _{\delta, -\nu -2} (M;TM)
\end{equation}

\noindent is surjective, and similarly
\begin{equation}\label{cvlinj}
\cvl : H^{k+2} _{-\delta,-\nu} (M;TM) \rightarrow H^k _{-\delta, -\nu -2} (M;TM)
\end{equation}

\noindent is injective for all $k\geq 0$. Furthermore, there exists a finite dimensional space $\YY$ consisting of smooth bounded vector fields for which
\begin{equation}\label{cvlsurj2}
\cvl : H^{k+2} _{-\delta,-\nu} (M;TM) \oplus \YY \rightarrow H^k _{-\delta, -\nu -2} (M;TM)
\end{equation}

\noindent is surjective.

\end{thm}

\noindent For our purposes, the most important of the three mapping properties enumerated by this theorem is the final one, and it will be useful to have the H\"older space version of this fact:

\begin{cor}\label{trumpetcvlcor}

Let $(M,g)$, $\delta$, and $\nu$ be as in Theorem \ref{trumpetcvlmap}. There exists a finite dimensional subspace $\YY'$ consisting of smooth bounded vector fields such that the map
\[ 
\cvl : C^{k+2,\alpha} _{-\delta,-\nu}(M;TM) \oplus \YY' \to C^{k,\alpha} _{-\delta,-\nu -2} (M;TM)
\]

\noindent is surjective.

\end{cor}

\noindent A justification of these results will be given in the appendix. We also note that the condition that $g$ admits no conformal Killing fields is known to be generic \cite{bcs}.

The last analytical tool we will need to prove Theorem \ref{mainthm} are global supersolutions and global subsolutions. Given a positive function $\phi$ on $M$ and a bounded generalized inverse $G$ for $\cvl$, let us denote by $W_\phi$ the vector field given by $G(b_n \phi^\kk \nabla \tau)$. Observe, in particular, that $W_0 = 0$. We define the operator $\lich_\phi (\theta)$ as
\begin{equation}\label{lichop}
\lich_\phi (\theta) = \Delta_g \theta - c_n R_g \theta - \beta \theta^{\kk-1} + c_n |\sigma + LW_\phi|^2 _g \theta ^{-\kk-1}.
\end{equation}

\noindent Before we define global barriers for the LCBY equations, it is important to note that the Lichnerowicz operator in \eqref{lichop} has the following useful property. If $u$ is a positive function on $M$, and we define a metric $\tilde g = u^{\kk-2}g$, then the Lichnerowicz operator $\tilde\lich_0$ associated to the metric $\tilde g$ is related to that of $g$ by
\begin{equation}\label{confcov}
\lich_0 (\theta) = u^{\kk-1}\tilde\lich_0 (u^{-1}\theta).
\end{equation}

\noindent We shall refer to this property as the \textit{conformal covariance of the Lichnerowicz equation}, and observe that an obvious consequence is that $\tilde \lich_0(u^{-1}\theta)$ and $\lich_0 (\theta)$ have the same sign at all points.

Given a pair of positive functions $\phi_-\leq \phi_+$, we shall call them a pair of \textit{global sub/supersolutions} of the system \eqref{lcby1}-\eqref{lcby2} if for every function $\phi$ satisfying $\phi_- \leq \phi \leq \phi_+$ we have $\lich_\phi (\phi_+) \leq 0$ and $\lich_\phi (\phi_-) \geq 0$. If these inequalities hold only in the weak sense, then they are said to be a pair of \textit{weak} global sub/supersolutions. The utility of weak global sub/supersolutions, a proof of which can be found in \cite{cm}, is the following fact.

\begin{prop}\label{subsupprop}
Let $(M,g)$ be a smooth Riemannian manifold and $F$ a locally Lipschitz function. Suppose that $\underline{\phi} \leq \overline{\phi}$ are continuous functions which satisfy
\[
\Delta_g \underline\phi \geq F(z,\underline\phi), \quad \Delta_g \overline\phi \leq F(z,\overline\phi)
\]

\noindent weakly at all points $z \in M$. Then there exists a smooth function $\tilde\phi$ on $M$ such that 
\[
\Delta_g \tilde\phi = F(z,\tilde\phi), \quad \underline\phi \leq \tilde\phi \leq \overline\phi.
\]
\end{prop}

\noindent In the following sections, we will drop all subscripts ``$g$'' from operators associated to the background metric.


\section{The case of positive scalar curvature}

\subsection{Global barriers}

 One can construct global sub/supersolutions for the system \eqref{lcby1}-\eqref{lcby2} once we have some control over the fall-off of $|LW_\phi|$ on the ends in terms of $\|\phi\|_0$. For this, we have the proposition

\begin{prop}\label{trumpetlbound}
Given a manifold $(M,g)$ which admits no bounded conformal Killing fields and conformal data $(g,\sigma,\tau)$ of trumpet type, then any solution $W_\phi$ of \eqref{lcby2} satisfies 
\begin{equation}\label{lbound2}
|LW_\phi| \leq K_\tau r^{-\gamma /2 -1}e^{-\delta x}\|\phi\|^\kk _0,
\end{equation}

\noindent where $K_\tau$ is a constant depending only on $\tau$ and the dimension $n$.
\end{prop}

\noindent The proof of this theorem is a straightforward generalization of the proof of \cite[Lemma 3.3]{lea2} and is left to the reader.

We proceed to construct global barriers for the system \eqref{lcby1}-\eqref{lcby2} with trumpet type background data first under the assumption that $R_g >0$ on $M$, and that $R_g \geq c > 0$ on $\ec$. In the following section we will show that, whenever the Yamabe invariant, defined as 
\[
Y(M,[g]) = \inf_{\substack{u\in C^\infty _0 \\ 0\leq u \not\equiv 0}} Q(u) \quad \mathrm{where} \quad Q(u) = \frac{\frac12 \int_M (|\nabla u|^2 + c_n Ru^2)}{\left( \int_M u^{\frac{2n}{n-2}} \right) ^{\frac{n-2}{n}}}
\]

\noindent  is positive, there exists a transformation of $g$ by a conformal factor with reasonable asymptotics which has positive scalar curvature everywhere. Straightforward modifications of the fixed point argument in the case where $R > 0$ then adapt to generate a solution to \eqref{lcby1}-\eqref{lcby2} under the weaker assumption that $Y(M,[g]) > 0$.

First note that since $g$ is of trumpet type, $R \rightarrow \mr R$ which is bounded away from zero on $\ec$, and this convergence is at the rate $e^{-x}$, and $R$ decays sufficiently rapidly on $\ef$ so that $\chi_e R \in L^2 _{-1,-\gamma -2}$. Just as in \cite{lea2}, we will construct global supersolutions first on both ends, and then show that for sufficiently small $\|\sigma\|_0$ we may find a constant which serves as an ``interpolating'' global supersolution. We first find a global supersolution on $\ef$ using an approach quite similar to that found in \cite{dimm}. Namely, first fix a positive smooth function $F$ which equals $r^{-\gamma-2}$ on $\ef$ and $e^{-x}$ on $\ec$. The positivity of $R$ makes clear that there is a unique positive solution $\Psi$ to
\begin{equation*}
(\Delta - c_n R)(\Psi + \chi_e) = -F.
\end{equation*}

\noindent Here we have assumed that $F$ is so large that $c_n \chi_e R - F < 0$ everywhere. Moreover, well-established regularity results for the conformal Laplacian $\LL = \Delta - c_n R$ on manifolds with AE ends (\cite{ciy}, \cite{dimm}) and on those with ends of cylindrical type \cite{cm} along with the local parametrix gluing principle described in \cite{cmp} imply that $\Psi$ has the decomposition $\Psi = c_\gamma r^{-\gamma} + \hat \Psi$, where $\hat\Psi \in C^{2,\alpha} _{-1,-2\gamma}$ and $c_\gamma$ is the constant $(\gamma^2 + (n-2)\gamma)^{-1}$. Note, in particular, that this means there is some $r_0 >1$ such that the function $1+\Psi$ is decreasing on $\ef$ wherever $r \geq r_0$. Fix such an $r_0$, and set $a = \min_{\{r=r_0\}}(1+\Psi) >1$.

We now show that, for sufficiently small $\eta > 0$, the function $\eta(1+\Psi)$ is a global supersolution on $\ef$. Just as in \cite{dimm}, we see that for any $0 < \phi \leq \eta (1+\Psi)$, we have
\begin{equation*}
\lich_\phi (\eta (1+\Psi)) = -\eta F - \beta (\eta(1+\Psi))^{\kk -1} + c_n|\sigma + LW_\phi|^2 (\eta(1+\Psi))^{-\kk -1}
\end{equation*}

\noindent which, using Proposition \ref{trumpetlbound} and standard estimates, is bounded above on $\ef$ by
\begin{equation*}
-\eta F + r^{-\gamma -2}(C_1 \eta^{\kk -1} + C_2 \|\sigma\|_0 ^2 \eta^{-\kk -1})
\end{equation*}

\noindent where the constants $C_1$ and $C_2$ depend only on $n$, $\tau$, and our fixed choice of $F$. Thus, recalling that $F = r^{-\gamma -2}$ on $\ef$, we may choose $\eta > 0$ so small that 
\begin{equation*}\label{etacond1}
-\frac{1}{2}F + C_1 \eta^{\kk -2} r^{-\gamma -2} < 0, 
\end{equation*}

\noindent and so for $\|\sigma\|_0$ sufficiently small,
\begin{equation*}
-\frac{1}{2}F + C_2 \eta^{-\kk -1}r^{-\gamma -2}\|\sigma\|^2 _0 < 0.
\end{equation*}

\noindent It is clear that such a choice for $\eta$ gives a global subsolution $\eta(1+\Psi)$ on all of $\ef$.

Next, to extend this global supersolution to a weak global supersolution on all of $M$, we argue as follows. Since $1+\Psi$ is decreasing for sufficiently large $r$, it is easy to see that there is some $r_1 > r_0$ such that 
\begin{equation*}
1 < \max_{r=r_1} (1+\Psi) < a.
\end{equation*}

\noindent Fix such an $r_1$, and observe that the positivity and asymptotics of $R$ means that there is some positive number $R_0$ such that $R \geq R_0$ on the manifold with boundary $M\setminus \{r\geq r_1\}$. Hence, we choose $\eta$ so small that \eqref{etacond1} is satisfied and $\eta < (R_0/4K_\tau ^2)^{1/(\kk -2)}$, and we then note that if $\sigma$ satisfies the additional smallness condition that the interval
\begin{equation*}
I = \left ( \left (\frac{\|\sigma\|_0}{2K_\tau}  \right )^{\frac{1}{\kk}}, \left (\frac{R_0}{4K_\tau ^2} \right )^{\frac{1}{\kk -2}} \right ) 
\end{equation*}

\noindent has non-empty intersection with $(\eta,a\eta)$, we may then choose some constant $\epsilon_+$ in this intersection. Just as in the case where $M$ admits only ends of cylindrical type, one easily checks that $\epsilon_+$ is a global supersolution on all of $M$ \cite{lea2}. Of course, the importance of choosing $r_0 < r_1$ as we have done is this guarantees that $\min\{\eta(1+\Psi),\epsilon_+\}$ is a weak global supersolution on $E_{r_0} := \ef \cap \{r\geq r_0\}$. Hence one may define a weak global supersolution $\phi_+$ to be this minimum on $E_{r_0}$, and to be identically $\epsilon_+$ on the rest of the manifold $M$. Now consider the equation \eqref{lcby1} with respect to the limiting metric on $\ec$, assume $W =0$ and suppose we are only considering solutions $\mr\phi$ which are functions on $\Sigma$ (in the ACC case) or $S^1 \times \Sigma$ (in the AP case). This equation can be written
\begin{equation}\label{redlich}
\mr \Delta \mr\phi - c_n \mr R \mr \phi - \mr\beta \mr\phi^{\kk-1} + c_n |\mr\sigma|_{\mr g}^2 \mr\phi^{-\kk-1} = 0,
\end{equation}

\noindent and we refer to this as the \textit{reduced Lichnerowicz equation}. It was proved in \cite{cm} that this equation has a unique positive solution, and in the computations which follow, we will assume that $\mr\phi$ is a smooth positive function equal to this solution on $\ec$ and to the constant $\eta$ on $\ef$.

Finally, localizing the discussion to the end $\ec$, we observe that we may thus find some sufficiently large constant $b$ such that $\min \{\epsilon_+, \mr\phi + bu\}$ is a weak global supersolution on $\ec$ where $u$ is a solution to the equation
\begin{equation*}
\Delta u - c_n R u = -\chi_c e^{-\nu x}
\end{equation*}

\noindent and $\nu > \delta/2\kk$. We thus obtain a weak global supersolution defined on all of $M$ which we continue to call $\phi_+$ by setting
\begin{equation*}
\phi_+ = \begin{cases} 
\min\{\eta(1+\Psi),\epsilon_+\} \quad &\mathrm{on} \quad \ef \\
\epsilon_+ \quad &\mathrm{on} \quad M\setminus (\ef \cup \ec) \\
\min\{\epsilon_+, \mr\phi + b e^{-\nu x}\} \quad &\mathrm{on} \quad \ec  
\end{cases}.
\end{equation*}

Now to construct a global subsolution, we begin by finding some $x_0 >1$ so large that
\[
\lich_0 (\mr\phi) + e^{-\nu x} - c_n |LW_\phi|(|LW_\phi| + 2|\sigma|)(\min_M \mr\phi)^{-\kk-1} \geq 0
\]

\noindent on $E_{x_0} = \ec \cap \{x\geq x_0\}$, and we then solve the elliptic boundary value problem
\begin{align*}
(\Delta - c_n R)v &= -e^{-\nu x} \\  v |_{\partial E_{x_0}} &=  \mr\phi  |_{\partial E_{x_0}}.
\end{align*}

\noindent It then follows that $\varphi = \mr\phi - v$ is a global subsolution wherever it is positive. To construct the rest of a global subsolution, we will proceed in a fashion similar to \cite{lea2}, and construct a positive weak global subsolution equal to $\varphi$ far out on $\ec$, and which asymptotically approaches $\eta$ from below on $\ef$. First, observe that on $\ef$ we have that, for any $C>0$,
\begin{equation}\label{euclap}
\Delta (1 -r^{-\nu}) = (\nu - \nu^2)r^{-\nu -2} + \bigo(r^{-\gamma -2}).
\end{equation}

\noindent Hence for any $\eta (1-Cr^{-\nu}) \leq \phi \leq \phi_+$, we have that $\lich_\phi (\eta(1-Cr^{-\nu}))$ is given by
\begin{multline*}
C \left [ (\nu - \nu^2)r^{-\nu-2} + \bigo(r^{-\gamma-2}) \right ] - c_n R \eta (1-Cr^{-\nu}) \\- \beta (\eta(1-Cr^{-\nu}))^{\kk -1} + c_n |\sigma + LW_\phi|^2 (\eta(1-Cr^{-\nu}))^{-\kk-1}
\end{multline*}

\noindent which is, of course, bounded below by
\begin{equation*}
C \left [ (\nu - \nu^2)r^{-\nu-2} + \bigo(r^{-\gamma-2}) \right ] - c_n R\eta -\beta\eta^{\kk -1} + c_n|\sigma + LW_\phi|^2 \eta^{-\kk -1}.
\end{equation*}

\noindent Thus given the known decay rates of the last three terms above (the final decay rate following from Proposition \ref{trumpetlbound}), we conclude that there exists some $r_2 > 1$ such that $\eta(1-Cr^{-\nu})$ is a global subsolution corresponding to $\phi_+$ for any choice of $C>0$. We may thus choose $C$ to be so large that this function is only positive where it is a global subsolution. 

Finally, we find our interpolating global subsolution as follows. We define $\KK$ to be the compact manifold with boundary given by
\begin{equation*}
\KK = \overline{M \setminus (\ef \cup \ec)}.
\end{equation*}

\noindent Obviously $\KK$ has a boundary with two connected components $\partial_1 \KK$ and $\partial_2 \KK$, the former lying in $\ef$ and being diffeomorphic to $S^{n-1}$ and the latter lying in $\ec$ being diffeomorphic to $\Sigma$. We then solve the following elliptic boundary value problem on $\KK$:
\begin{align*}
(\Delta - c_n R - \beta)w &= 0 \\ w|_{\partial_1 \KK} &= 0 \\ w|_{\partial_2 \KK} &= \frac{1}{2}\mr\phi|_{\partial_2 \KK}.
\end{align*}

\noindent We easily see that $w$ is positive on the interior of $\KK$ by the strong maximum principle, and that $\eta(1-Cr^{-\nu}) > w$ on $\partial_1 \KK$ and $\varphi > w$ on $\partial_2 \KK$. Hence we may define our weak global subsolution as
\begin{equation}\label{trumpetglobsub}
\phi_- = \begin{cases} \max\{w,\eta(1-Cr^{-\nu})\} \quad &\mathrm{on} \quad \ef  \\ w \quad &\mathrm{on} \quad M\setminus (\ef \cup \ec) \\ \max\{\varphi,w\} \quad &\mathrm{on} \quad \ec \end{cases}.
\end{equation}

\noindent We end this section by emphasizing that these barrier constructions generalize to the case of multiple ends which are AE or of cylindrical type in an obvious way.

\subsection{A fixed point argument}

We must first verify that the solution $\tilde\phi$ we obtain from Proposition \ref{subsupprop} for every $\phi_- \leq \phi \leq \phi_+$ is unique. However, the fact that both $\phi_-$ and $\phi_+$ approach the same limiting function asymptotically on both ends means one can apply an obvious extension of \cite[Proposition 3.2]{lea2} to trumpet data whose proof is left to the reader. Next, extend $\mr\phi$ to a smooth positive function on all of $M$ which is identically $\eta$ on $\ef$ and bounded between $\phi_-$ and $\phi_+$, and fix a bounded generalized inverse function 
\begin{equation*}
G : C^{0,\alpha} _{-\delta, -\gamma/2 -2} (M; TM) \rightarrow C^{2,\alpha} _{-\delta,-\gamma/2}(M;TM) \oplus \YY
\end{equation*}

\noindent for $\cvl$. We may choose some positive exponents $\nu' < \nu$ and $\mu' < \gamma/2$, and we define a closed, convex subset of the Banach space $C^0 _{-\nu',-\mu'}(M)$ as
\begin{equation*}
U = \{ \psi \in C^{0} _{-\nu',-\mu'}(M) : \phi_- -\mr\phi \leq \psi \leq \phi_+ - \mr\phi \},
\end{equation*} 

\noindent and on this subset we define the map $\WW : \psi \mapsto G(b_n(\mr\phi + \psi)^\kk \nabla\tau)$. We may also define the map $\sss : \mathrm{ran} (L \circ \WW) \rightarrow C^{2,\alpha} _{-\nu,-\gamma}(M)$ by $\sss (\pi) = \QQ(\sigma + \pi) - \QQ(\sigma)$, where again $\QQ$ is the unique solution of the Lichnerowicz equation
\begin{equation*}
\Delta \theta - c_n R \theta - \beta \theta^{\kk -1} + c_n |\sigma + \pi|^2 \theta^{-\kk -1} =0
\end{equation*}

\noindent guaranteed by \cite[Theorem 6.1]{cm}. 

Just as in \cite{lea2}, we will use the following fixed point theorem:

\begin{thm}\label{fixedpt}
Let $\mathcal{B}$ be a Banach space, and let $U \subset \mathcal{B}$ be a non-empty, convex, closed, bounded subset. If $\mathcal T : U \to U$ is a compact operator, then there exists a fixed point $z \in U$ such that $\mathcal T (z) = z$.
\end{thm}

\noindent A proof of this theorem can be found in \cite{ist}, and we shall apply it to the composition of $\sss \circ L \circ \mathcal W$ with the compact embedding $C^{2,\alpha} _{-\nu,-\gamma} \hookrightarrow C^0 _{-\nu',-\mu'}$ to obtain a solution to the LCBY equations for trumpet-type conformal data as a fixed point of this composition. Since this composition is obviously a compact mapping from the closed, convex set $U$ to itself, we need only verify that this map is continuous. All of its constituent maps are clearly continuous except for $\sss$, and so we need the following lemma.

\begin{lem}\label{accont}
If $R > 0$, then the map $\sss : \mathrm{ran}(L\circ \mathcal W) \to C^{2,\alpha} _{-\nu,-\gamma}$ is continuous.
\end{lem}

\proof This follows from the implicit function theorem (see \cite{sid}, for example). We define a function $F : C^{2,\alpha} _{-\nu,-\gamma} \times C^{2,\alpha} _{-\delta, -\gamma/2 -1} \to C^{0,\alpha} _{-\nu, -\gamma-2}$ by
\[
F(\psi,\pi) = (\Delta - c_n R)(\mr\phi + \psi) - \beta (\mr\phi + \psi)^{\kk-1} + c_n |\sigma + \pi|^2 (\mr\phi +\psi)^{-\kk-1}.
\]

\noindent Observe that $F(\sss(\pi),\pi) = 0$ by definition. The Fr\'{e}chet derivative $DF_{(\psi,\pi)} (h,k)$ of $F$ at the point $(\psi,\pi)$ acting on pairs $(h,k) \in C^{2,\alpha} _{-\nu,-\gamma} \times C^{2,\alpha} _{-\delta, -\gamma/2 -1}$ is given by
\begin{multline*}
(\Delta - c_n R)h - (\kk-1)\beta (\mr\phi + \psi)^{\kk-2} h + 2c_n \langle \sigma + \pi, k \rangle (\mr\phi + \psi)^{-\kk-1} \\ - c_n (\kk+1)|\sigma + \pi|^2 (\mr\phi + \psi)^{-\kk-2} h.
\end{multline*}

\noindent Hence we may write 
\[
DF_{(\psi,\pi)}(h,0) = (\Delta - H)h
\]

\noindent where 
\[
H = c_n R + (\kk-1)\beta (\mr\phi + \psi)^{\kk-2}  + c_n (\kk+1)|\sigma + \pi|^2 (\mr\phi + \psi)^{-\kk-2}.
\]

\noindent Therefore, given any pair $(\psi,\pi)$ for which $\mr\phi + \psi >0$ everywhere, which is certainly true for all $\psi$ in the image of $\sss$, we see that $DF_{(\psi,\pi)}(\cdot,0) : C^{2,\alpha} _{-\nu, -\gamma} \to C^{0,\alpha} _{-\nu,-\gamma-2}$ is an isomorphism. Since it is obviously continuous in $(\psi,\pi)$, the implicit function theorem implies $\sss$ is continuous in an open neighborhood of $\pi$. \qed

\vspace{.5cm}

\noindent Having proved the lemma, Theorem \ref{fixedpt} gives us a solution to the system \eqref{lcby1}-\eqref{lcby2} for trumpet data in the case where $R > 0$. This proves Theorem \ref{mainthm} in the case of trumpet conformal data with positive scalar curvature.


\subsection{Yamabe positivity}

Consider again the extreme Kerr metric mentioned in the introduction. The $t$-constant slices of this spacetime in Boyer-Lindquist coordinates do not have pointwise positive scalar curvature, though one may show that these trumpet-type slices do, in fact, have positive Yamabe invariant \cite{ad}. We would thus like to extend our analysis above to the case where we know only that $Y(M,[g]) > 0$. We begin by proving that this condition implies the existence of a conformal factor which transforms a given Yamabe positive metric to one with pointwise scalar curvature. To prove this, we first need the following lemma.

\begin{lem}\label{trumpposscal1}
Let $(M,g)$ by a Riemannian manifold with one asymptotically Euclidean end and one end which is asymptotically cylindrical or periodic, and suppose further that $\chi_e(g - g_{Euc}) + \chi_c(g - \mr g) \in C^{2,\alpha} _{-1,-\gamma}$. Then if $Y(M,[g]) > 0$, the first eigenvalue of the operator 
\begin{equation*}
-\mr \LL = -\mr \Delta _g + c_n \mr R,
\end{equation*}

\noindent defined on the model cylindrical or periodic metric of $\ec$, is positive.
\end{lem}

\proof Our proof is a slight adaptation of those for Lemmas 2.7 and 2.9 in \cite{ab}. Let $\lambda_0$ be the first (i.e., smallest) eigenvalue of $-\mr\LL$. There is thus a positive function $\varphi_0 \in C^\infty (\Sigma)$ with $\|\varphi_0\|_{L^2(\Sigma)} = 1$ which satisfies $-\mr\LL \varphi_0 = \lambda_0 \varphi_0$. We then define a Lipschitz function $v_\vep$ whose domain is $\ec$ and depends only on $x$. This function is supported in the region $1+\frac{1}{\vep} \leq x \leq 3 + \frac{2}{\vep}$, is identically equal to $\vep$ where $2+\frac{1}{\vep} \leq x \leq 2 + \frac{2}{\vep}$, and has $|v_\vep '| = \vep$ wherever this derivative is supported.

\begin{figure}[h!]
\centering
\begin{tikzpicture}

\draw (0,-.25) -- (0,2);
\draw (-.25,0) -- (13,0);

\draw (1,-.1) node [below] {$1 + \frac1\vep$} -- (1,.1);
\draw (3,-.1) node [below] {$2 + \frac1\vep$} -- (3,.1);
\draw (10,-.1) node [below] {$2 + \frac2\vep$} -- (10,.1);
\draw (12,-.1) node [below] {$3 + \frac2\vep$} -- (12,.1);
\draw (-.1,1) node [left] {$\vep$} -- (.1,1);

\draw (1,0) -- (3,1) -- (10,1) node [above,midway] {$v_\vep (x)$} -- (12,0);

\end{tikzpicture}
\caption{The graph of the function $v_\vep$.}
\end{figure}

Next, in the case of one ACC end, we define a test function $w_\vep(x,y) = v_\vep (x) \varphi_0 (y)$ (this definition makes sense since the function $v_\vep$ effectively localizes to the end of cylindrical type). Consider the Yamabe quotient of this function, given by
\begin{align*}
Q(w_\vep) &= \frac{\frac12 \int_M \left ( |\nabla w_\vep |^2 + c_n R w_\vep ^2 \right )}{\left ( \int_M w_\vep ^{\frac{2n}{n-2}}  \right )^{\frac{n-2}{n}}} \\
&= \frac{\frac12 \int_M \left ( \left |v_\vep ' \varphi_0 \partial_x + v_\vep \mr{\nabla} \varphi_0 + \bigo (e^{-x}) \right |^2  +c_n \mr R w^2 _\vep + \bigo (e^{-x} ) \right )}{\left ( \int^\infty _0 v_\vep^{\frac{2n}{n-2}}(x) dx \right )^{\frac{n-2}{n}} \left ( \int_\Sigma \varphi_0 ^{\frac{2n}{n-2}}(y) dy \right )^{\frac{n-2}{n}}} \\ 
&\leq \frac{\frac12 \left ( 2\vep^2 \int_\Sigma \varphi_0 ^2  + \vep \lambda_0 \int_\Sigma \varphi_0 ^2 \right ) + \bigo (e^{-\frac1\vep})}{\left ( \frac1\vep \cdot \vep^{\frac{2n}{n-2}}\cdot C \right )^{\frac{n-2}{n}}} \\
&\leq c_1 \vep^{n-2}{n} + c_2 \lambda_0 \vep^{-\frac2n} + c_3 e^{-\frac1\vep}.
\end{align*}

\noindent We thus see that $Q(w_\vep) \to -\infty$ as $\vep \to 0$ if $\lambda_0 < 0$, which implies $Y(M,[g]) = -\infty$, a contradiction. Similarly, $Q(w_\vep) \to 0$ as $\vep \to 0$ if $\lambda = 0$ which implies $Y(M,[g]) \leq 0$. It thus follows that, since $Y(M,[g]) > 0$ by assumption, we must have $\lambda_0 > 0$, proving the assertion in the case of one ACC end. If the end of cylindrical type is AP, we may just define $w_\vep = v_\vep (x) \varphi_0 (x,y)$ where $\varphi_0$ is the lift to the cylinder of a simple eigenfunction for $-\mr{\mathcal{L}}$ on $S^1 \times \Sigma$. The estimate above for the Yamabe quotient is even simpler in this case and left to the reader. \qed

\vspace{1pc}

\noindent Using this lemma, we first show that we can conformally transform a Yamabe positive metric to one whose scalar curvature is positive and bounded away from zero sufficiently far out on $\ec$. From the lemma, we know that there exists some positive function $\mr\vartheta \in C^\infty (\Sigma)$ such that $\mr\LL \mr\vartheta = -\lambda_0 \mr\vartheta$ where $\lambda_0 > 0$. Then if we extend this function to a smooth positive function which is identically $1$ outside of a neighborhood of $\ec$, and define a new metric $\tilde g = \mr\vartheta^{\kk -2} g$, we find that on $\ec$,
\begin{equation*}
\LL \mr\vartheta = -c_n \mr\vartheta^{\kk -1} \tilde R.
\end{equation*}

\noindent Since the left side of this equation asymptotically approaches $-\lambda_0 \mr\vartheta$, we conclude that $\tilde R \geq c > 0$ for sufficiently large $x$. Suppose this is true where $x \geq x_0$. This lemma will also allow us to show that the Yamabe positivity of the metric will permit us to prescribe positive scalar curvature (with appropriate fall-off on each end) on all of $M$. 

\begin{prop}\label{trumpposscal2}
Let $(M,g)$ satisfy the hypotheses of Lemma \ref{trumpposscal1}. Then there exists a positive function $u$ on $M$ which satisfies $u \rightarrow 1$ on $\ef$ and $u \rightarrow \mr\vartheta$ on $\ec$ such that, if $\tilde g = u^{\kk -2}g$, then $\tilde R > 0$ everywhere. Furthermore, $\tilde R$ is bounded away from zero on $\ec$.
\end{prop}

\proof We first choose $\mr \vartheta$ as in the preceding paragraph and continue to call the conformally transformed metric $g$, so that $R \geq c > 0$ where $x \geq x_0$. We then look for a conformal factor of the form $u = 1+v$ where $v \in H^2 _{-\nu,-\mu}$, where $0 < \nu < \nu_0$ and $\mu < \frac{1}{2}$, such that $\LL u = f$. Here $\nu_0$ is chosen so that the operator $\LL$ has no indicial roots with imaginary part in the range $[-\nu_0,\nu_0]$, and $f$ is a strictly negative function which identically equals $-c_n R$ for $x \geq x_0$, and is bounded below by $-Cr^{-\gamma -2}$ for some positive $C$. First observe that if we can show
\begin{equation*}
\LL u = f \Longleftrightarrow \LL v = f + c_n R =: \tilde f,
\end{equation*} 

\noindent where $u$ is positive, then our proof will be complete by the standard identity $\LL u= -c_n u^{\kk -2}\tilde R$. We would thus like to show that $\tilde f$ lies in the image of the map 
\begin{equation*}
\LL : H^2 _{-\nu, -\mu}(M) \rightarrow L^2 _{-\nu, -\mu -2}(M).
\end{equation*}

\noindent Since constructions of parametrices for $\LL$ are well known for both ends of cylindrical type and for asymptotically Euclidean ends, standard arguments (see \cite[Appendix C]{cmp}) show that this map is Fredholm. Elliptic regularity thus implies that, to show that this map is surjective, it suffices to show that $\LL^* = \LL$ is injective as a map
\begin{equation*}
\LL : H^2 _{\nu,2-n+\mu} (M) \rightarrow L^2 _{\nu, -n + \mu}(M).
\end{equation*}


Observe that our choice of $\nu < \nu_0$ means that any element in the null space of this map also lies in $H^2 _{-\nu,2-n+\mu}$. Hence, we need only to check the injectivity of $\LL$ as a map
\begin{equation*}
\LL : H^2 _{-\nu,2-n+\mu}(M) \rightarrow L^2 _{-\nu,-n+\mu}(M).
\end{equation*}

\noindent Next, since $R$ decays like $r^{-\gamma -2}$, the indicial roots coming from this operator's restriction to $\ef$ are the same as those of the Laplacian, and hence the decay of any solution to $\LL w = 0$ can be no slower than that corresponding to the first indicial root with imaginary part below $2-n+\mu$, which is $2-n$. This means we may, in fact, assume that $w \in H^{2} _{-\nu,2-n}$. Given such a nontrivial solution, it follows from the definition of this function space that $\nabla w \in H^1 _{-\nu,1-n}$, and since $1-n < -n/2$ for all $n \geq 3$, we have $\nabla w \in L^2$ and therefore
\begin{equation*}
\int_M |\nabla w|^2 < \infty.
\end{equation*}

\noindent Moreover, observe that since $\chi_e R \in C^{0,\alpha} _{-1,-\gamma -2}$, there exists some positive constant $C$ such that $R \leq C r^{-\gamma -2}$. But of course $w \in H^2 _{-\nu,2-n}$ implies that
\begin{equation*}
\int_M r^{n-4}w^2 < \infty,
\end{equation*}

\noindent and therefore since $n-4 > -\gamma -2$, we are also able to conclude that
\begin{equation*}
\int_M R w^2 < \infty.
\end{equation*} 

Finally, it follows from standard Sobolev embedding arguments \cite{bar} that for any $\rho \in \RR$, we have
\begin{equation*}
\|w\|_{0,2n/(n-2),-\nu,\rho} \leq C \|w\|_{1,2,-\nu,\rho}
\end{equation*}

\noindent for some $C>0$. But since we clearly have $\|w\|_{0,q,-\nu,\rho'} \leq \|w\|_{0,q,-\nu,\rho}$ whenever $\rho' \geq \rho$, and since 
\begin{equation*}
-\frac{n}{2^*} = \frac{2-n}{2} \geq 2-n,
\end{equation*}

\noindent we conclude that $w \in L^{2^*}(M)$. It thus follows that the quantity
\begin{equation*}
\frac{\frac{1}{2}\int_M (|\nabla w|^2 + Rw^2)}{\left ( \int_M w^{\frac{2n}{n-2}} \right )^{\frac{n-2}{n}}}
\end{equation*}

\noindent is finite. Since $w$ is also smooth, one may cut off $w$ to be supported in larger and larger compact sets to obtain a minimizing sequence for the Yamabe functional. However, since $\LL w = 0$, this sequence minimizes the functional to zero, and so $Y(M,[g]) = 0$, a contradiction. Hence we may find a solution of the equation $\LL v = \tilde f$ where $v \in H^2 _{-\nu,-\mu} (M)$.

We thus need only show that $u = 1 + v$ is positive to prove the proposition. For this we employ an argument used in \cite{cm}. First, we clearly have that $u$ is positive and bounded away from zero for sufficiently large $x$ and $r$, so we choose a compact set with smooth boundary which we will again call $\KK$ such that $u > 1/2$ away from $\KK$. It follows from the strict monotonicity of the Yamabe functional that $Y(\KK,[g]) > 0$, and it follows from the existence of a Sobolev inequality on $\KK$ (see \cite[Lem. 4.1]{cm}, for example) that the first Dirichlet eigenvalue of $-\LL$ on $\KK$ is positive. Call this eigenvalue $\lambda_0 (\KK)$, and let $\phi_0$ be a corresponding positive eigenfunction. Next, by writing $\LL [\phi_0 (u/\phi_0)] = f$ on the interior of $\KK$, we have that
\begin{equation}\label{ineq4}
\Delta \left ( \frac{u}{\phi_0} \right ) + 2 \left\langle \frac{\nabla \phi_0}{\phi_0} , \nabla \left ( \frac{u}{\phi_0} \right )\right\rangle - \frac{\lambda_0(\KK)}{\phi_0} \frac{u}{\phi_0} = \frac{f}{\phi_0} < 0.
\end{equation}

\noindent Now the function $u/\phi_0$ is negative precisely where $u$ is, and since $u/\phi_0 \rightarrow +\infty$ as one approaches the boundary of $\KK$, if $u/\phi_0 < 0$ anywhere then this function attains a negative minimum. However, the existence of a negative minimum clearly contradicts \eqref{ineq4}, so this contradiction proves that $u > 0$ everywhere, and hence the proposition. \qed 

\vspace{.5cm}

To prove Theorem \ref{mainthm} in general, we first construct global barriers for the LCBY equations with respect to a background metric with one ACC end and one AE end. For this we take a much simpler approach than in \cite{lea2}. Having constructed these barriers, the proof of Theorem \ref{mainthm} will run just as in the positive scalar curvature case, though we will need to modify our proof that the solution operator $\sss : \pi \mapsto \QQ(\sigma + \pi) - \mr\phi$ is continuous. This is because the proof of Lemma \ref{accont} depends on the positivity of scalar curvature. 

\begin{prop}\label{confacbarriers}
Given conformal data $(g,\tau,\sigma)$ of trumpet type for which $Y(M,[g]) > 0$, there exists a smooth positive function $\mr \phi$ and a pair of global barriers $\phi_- \leq \phi_+$ for the LCBY equations such that $\phi_\pm \to \mr\phi_\ell$ on each end as $x \to \infty$.
\end{prop}

\proof We first need to address the existence of a positive solution to the reduced Lichnerowicz equation in the ACC case. Given any metric $g$ and a conformally related metric $\tilde g = u^{\kk -2}g$, there is a well-known identity expressing the Laplacian $\tilde \Delta$ associated to $\tilde g$ in terms of that of $g$, namely
\begin{equation*}
\tilde \Delta \phi = u^{2-\kk} (\Delta \phi + 2 \langle \nabla \log u, \nabla \phi \rangle_g ).
\end{equation*}

\noindent Now if we assume that $g = u^{\kk-2}(dx^2 + \mr g + \bigo(e^{-\omega x}))$ and let $u \to \mr u$, which is the case in which we are interested, the reduced Lichnerowicz equation we would like to solve is
\begin{equation}\label{confredlich}
\mr u^{2-\kk}(\mr\Delta \mr\phi + 2 \langle \mr\nabla \log \mr u, \mr\nabla \mr\phi \rangle_{\mr g}) - c_n \mr R \mr\phi -\beta \mr\phi^{\kk-1} + c_n \mr\sigma^2 \mr\phi^{-\kk -1} = 0.
\end{equation}

\noindent By conformal covariance and Proposition \ref{trumpposscal2}, there is no loss of generality in assuming that $\mr R > 0$ (to be more explicit, if $\vt$ is as in Proposition \ref{trumpposscal2}, and $\mr\phi$ is a translation invariant solution of the reduced equation with $\mr{\tilde R} > 0$, then conformal covariance of the Lichnerowicz equation on the cylindrical end implies that $\mr\vt \mr\phi$ is a solution to the original reduced equation). We thus see that any sufficiently large positive constant is a supersolution for the operator on the right hand side of \eqref{confredlich}. To construct a subsolution, we first claim that we may solve the equation
\begin{equation}\label{linconfredlich}
\mr u ^{2-\kk}(\mr \Delta \mr v + 2 \langle \mr\nabla \log \mr u, \mr\nabla \mr v \rangle_{\mr g} ) - (c_n \mr R + \mr \beta)\mr v = -c_n \mr\sigma ^2.
\end{equation}

\noindent It is easy to check that $\rho \mr v$ is a positive subsolution for the operator in \eqref{confredlich} for any sufficiently small $\rho > 0$. To see that \eqref{confredlich} has a solution, define a one-parameter family of Fredholm operators $P_t : H^{k+2}(\Sigma) \to H^k (\Sigma)$ by
\begin{equation}
P_t = \mr u _t ^{2-\kk} (\mr \Delta + 2t \langle \mr\nabla \log \mr u, \mr\nabla \cdot \rangle_{\mr g}) - (c_n \mr R + \mr \beta), \quad \mr u _t = 1 - t + t\mr u.
\end{equation}

\noindent Clearly $P_0$ has index zero, and so the local constancy of the Fredholm index implies that $P_1$ fails to be surjective if and only if there is a nontrivial solution to $P_1 \mr w = 0$. But the maximum principle prevents this equation from having any nontrivial solutions, so we thus conclude that \eqref{linconfredlich} has a solution which is nontrivial since $\mr\sigma^2 \not\equiv 0$, and is positive by the maximum principle. Since we have constructed positive barriers for the reduced Lichnerowicz equation, we are thus guaranteed a positive solution to this equation by the monotone iteration scheme. 

Let $\vartheta$ be a function described in Proposition \ref{trumpposscal2}, so that the metric $\tilde g = \vartheta^{\kk -2}g$ satisfies $\tilde R > 0$. Since $\lich_0 (\phi)$ and $\widetilde\lich_0 (\vartheta^{-1}\phi)$ have the same sign everywhere, it thus follows that $\phi_+$ is a global supersolution of the LCBY equations if and only if $u_+ = \vartheta^{-1} \phi_+$ is a global supersolution of the auxiliary system 
\begin{align}
(\tilde\Delta - c_n \tilde R )u - \beta u^{\kk -1} &= c_n u^{-\kk -1} |\tilde \sigma + \vartheta^{-2} LW|^2 _{\tilde g} \label{aux1} \\
\cvl W &= \frac{n-1}{n}\vartheta^\kk u^\kk \nabla\tau. \label{aux2}
\end{align}

\noindent Here, of course, we have defined $\tilde\sigma = \vartheta^{-2}\sigma$. Similarly, $\phi_-$ is a corresponding global subsolution of the LCBY equations if and only if $u_- = \vartheta^{-1}\phi_-$ is a global subsolution of \eqref{aux1}-\eqref{aux2}. Observe now that, in light of Proposition \ref{trumpetlbound}, the coefficients in this auxiliary system have precisely the same signs and decay properties as did the LCBY system with positive scalar curvature. As we know we can find some positive solution $\mr u$ to the corresponding reduced equation, one argues just as in the positive scalar curvature case to construct a sub/supersolution pair with asymptotics $u_\pm \to \mr u$ on $\ec$ and $u_\pm \to 1$ on $\ef$. Then if we define $\mr \phi = \mr\vartheta \mr u$, and set $\phi_\pm = \vartheta u_\pm$, the proposition is proved.\qed

\vspace{.5cm}

Having constructed a pair of global sub/supersolutions with the correct asymptotic limits, we may apply a fixed point argument similar to that for the positive scalar curvature case to find a solution to \eqref{lcby1}-\eqref{lcby2}. For this we define the set $U$ and the maps $\mathcal{W}$ and $\mathcal{Q}$ just as above. We note that $\lich_0 (\mathcal{Q}(\sigma)) = 0$, and so we simply replace ``$\mr{\phi}$'' with ``$\mathcal{Q}(\sigma)$'' in the definition of the map $\mathcal{S}_\sigma (\pi)$. The advantage of this replacement is that the conformal covariance of the Lichnerowicz equation gives us the identity
\begin{align}
\mcs_\sigma (\pi) = \QQ(\sigma + \pi) - \QQ(\sigma) &= \theta \hat{\QQ}(\theta^{-2} (\sigma + \pi)) - \theta \hat{\QQ}(\theta^{-2} \sigma) \label{id1} \\
&:= \theta \hat{\mcs}_{\theta^{-2} \sigma} (\theta^{-2} \pi) \label{id2}
\end{align}

\noindent where $\hat{\QQ}$ is the solution operator for the Lichnerowicz equation with respect to the metric $\theta^{\kk-2}g$. 

The proof of Theorem \ref{mainthm} in general reads exactly as the proof of the theorem in the positive scalar curvature case once we have established the analogue of Lemma \ref{accont}. For this we use an implicit function theorem argument very similar to the proof of that lemma, but we must deal with the fact that we no longer have $R > 0$. We shall get around this difficulty using a trick used by Maxwell in \cite{max1}, which essentially boils down to the identity \eqref{id1}.

\begin{lem}\label{confaccont}
If $Y(M,[g]) >0$, the map $\mathcal{S}_\sigma : \ran(L\circ \mathcal{W}) \rightarrow C^{2,\alpha} _{-\nu}(M)$ is continuous.
\end{lem}

\proof Given $\pi_0 \in C^{0,\alpha} _{-\delta, -\gamma/2-1} (S^2 _0 (M))$, we set $\theta_0 = \QQ(\sigma + \pi_0)$. If we also set $\hat{\sigma} = \theta_0 ^{-2} \sigma$ and $\hat{\pi}_0 = \theta_0 ^{-2} \pi_0$, and define $\hat{\mcs}_{\hat{\sigma}}$ to be the expression in \eqref{id2}, we have $\hat{\mcs}_{\hat{\sigma}} (\hat{\pi}_0) = 1 - \hat{\QQ}(\hat{\sigma})$. We next define the map $F : C^{2,\alpha} _{-\nu,-\gamma}(M) \times C^{0,\alpha} _{-\delta,-\gamma/2-1}(S^2 _0 (M)) \rightarrow C^{0,\alpha} _{-\nu,-\gamma-2}(M)$ by 
\begin{equation*}
F(\psi,\pi) = (\hat{\Delta} - c_n \hat{R})(\hat{\QQ}(\hat{\sigma}) + \psi) - \beta (\hat{\QQ}(\hat{\sigma}) + \psi)^{\kk -1} + c_n |\sigma + \pi|^2 (\hat{\QQ}(\hat{\sigma}) + \psi)^{-\kk -1}.
\end{equation*}

\noindent The Fr\'{e}chet derivative is computed exactly as in the proof of Lemma \ref{accont}, and we find that at the point $(\hat{\psi}_0, \hat{\pi}_0) = (\hat{\mcs}_{\hat{\sigma}}(\hat{\pi}_0),\hat{\pi}_0) = (1 - \hat{\QQ}(\hat{\sigma}),\hat{\pi}_0)$, we have
\begin{equation*}
DF_{(\hat{\psi}_0,\hat{\pi}_0)} (h,0) = [ \hat{\Delta} - (c_n \hat{R} + (\kk -1) \beta + c_n (\kk +1) |\hat{\sigma} + \hat{\pi}_0|^2)]h.
\end{equation*}

\noindent However, as $\hat{\QQ}(\hat{\sigma}+\hat{\pi}_0) = 1$, we have $-c_n \hat{R} = \beta - c_n |\hat{\sigma}+\hat{\pi}_0|^2$, so that in fact
\begin{equation*}
DF_{(\hat{\psi}_0,\hat{\pi}_0)} (h,0) = \left [ \hat{\Delta} - \left ((\kk -2) \beta + c_n (\kk +2) |\hat{\sigma} + \hat{\pi}_0|^2 \right ) \right ]h.
\end{equation*}

\noindent The map $DF_{(\hat{\psi}_0,\hat\pi_0)}(\cdot,0) : C^{2,\alpha} _{-\nu,-\gamma} (M) \rightarrow C^{0,\alpha} _{-\nu,-\gamma-2} (M) $ is thus an isomorphism, and so we conclude by the implicit function theorem that $\hat{\mcs}_{\hat{\sigma}}$ is continuous. The lemma now follows from (\ref{id1}) and (\ref{id2}), since we have $\mcs_\sigma (\pi) = \theta_0 \hat{\mcs}_{\hat{\sigma}} (\theta_0 ^{-2} \pi)$ for all $\pi \in C^{0,\alpha} _{-\delta, -\gamma/2-1} (S^2 _0 (M))$. \qed


\section*{Acknowledgment} This paper was completed while the author was in residence at the Mathematical Sciences Research Institute in Berkeley, California during the Spring 2016 semester, supported by the NSF Grant No. DMS-1440140. The author thanks Rafe Mazzeo for many helpful discussions.


\appendix

\section{The conformal vector Laplacian}\label{app}

In this appendix, we shall discuss the proofs of Theorem \ref{trumpetcvlmap} and Corollary \ref{trumpetcvlcor}. Detailed proofs can be found in \cite{lea2}, and while they draw on mostly standard analytical results, they are somewhat long and so we do not include them in their entirety here. 

\subsection{The Sobolev case}

The proof of a theorem nearly identical to Theorem \ref{trumpetcvlmap} can be found in \cite{cmp} in the case where the manifold has no AE ends, but their proof extends to the trumpet case in an obvious way. However, the result we need requires the condition of no \textit{bounded} conformal Kiling fields, for otherwise Proposition \ref{trumpetlbound} is not known to hold. To be precise, a straightforward generalization of the argument in \cite{cmp} proves that the map \eqref{cvlsurj1} is surjective. This surjectivity then allows us to quote \cite[Theorem 7.14]{maz} (in the ACC case) or an analogue of \cite[Lemma 4.18]{mpu} for the conformal vector Laplacian (in the AP case) to show that if $\cvl Y = W \in H^k _{-\delta,-\nu-2}$, then $Y \in H^k _{\delta,-\nu} $ must have a decomposition $Y = Y' + \mr Y_b + \mr Y_l$, where $Y'$ is a vector field in $H^{k+2} _{-\delta}$, and $\mr Y_b + \mr Y_l \in \YY_b \oplus \YY_l$. Here $\YY_b$ is a finite-dimensional space of bounded vector fields which satisfy $\mrcvl Y = 0$ far out on the end of cylindrical type, and $\YY_l$ is a finite-dimensional space of vector fields with linearly growing magnitude on each end of cylindrical type which also satisfy $\mrcvl Y = 0$. 

We note that $\YY_b$ is precisely the space $\YY$ defined in Theorem \ref{trumpetcvlmap}, and that \cite[Corollary 5.4]{cmp} guarantees that we may always take $\mr Y_b = \mr Y_l = 0$ in the AP case since we are assuming that the manifold admits no bounded conformal Killing fields. However, the reason why we may take $\mr Y_l = 0$ in the ACC case was not carefully explained in that paper. In fact, the H\"older space version of \cite[Theorem 6.1]{cmp} is false if one only assumes no $L^2$ conformal Killing fields (see \cite[Section 2.5]{lea1}), and so we clarify this point for ACC ends here. 

It is easy to see from the expression for $Y$ given in \cite[Thm. 7.14]{maz} that $k := \dim \YY_b = \dim \YY_l$, and in fact Melrose's Relative Index Theorem \cite[Thm. 6.5]{Mel} implies that the kernel of $\cvl$ is a $k$-dimensional subspace of $\YY_b \oplus \YY_l$. We claim our assumption of no bounded conformal Killing fields implies that $\ker (\cvl)\cap \YY_b = 0$, which, by linear algebra, means we can simply subtract off elements of $\ker (\cvl)$ from our expansion for the vector field $Y$ to obtain some $Y \in H^{k+2} _{-\delta} \oplus \YY_b$ which solves $\cvl Y = W$. 

To see this, suppose $V \in \ker (\cvl) \cap \YY_b$. Letting $M_A = M \cap \{x \leq A\}$, we proceed to integrate the equation $\langle V, \cvl V \rangle _g = 0$ over the manifold with boundary $M_A$:
\begin{equation*}
0 = \int_{M_A} \langle V, \cvl V \rangle _g = -\int_{M_A} |LV|^2 + \int_{\partial M_A} V_i (LV)^{0i}.
\end{equation*}

\noindent But since $V \in \YY_b$, the integrand of the boundary integral decays exponentially, namely like $e^{-\omega A}$. Hence taking the limit $A\to \infty$ gives us 
\begin{equation*}
\int_M |LV|^2 = 0
\end{equation*}

\noindent which contradicts our assumption that we have no bounded conformal Killing fields. This shows that we may take $\YY = \YY_b$ in the statement of Theorem \ref{trumpetcvlmap}, and so the theorem is proved. \qed


\subsection{The H\"older case}

Now consider $W \in C^{k,\alpha} _{-\delta,-\nu-2} (M;TM)$. Since for any $0< \delta' < \delta$ we have $W \in H^k _{-\delta',-\nu-2}(M;TM)$, we may quote Theorem \ref{trumpetcvlmap} to show that there is some $Y \in H^{k+2} _{-\delta',-\nu} \oplus \YY$ such that $\cvl Y = W$. However, in the ACC case, the mapping properties of the parametrices constructed in \cite{maz} immediately imply that $Y \in C^{k+2}_{-\delta,-\nu}(M;TM) \oplus \YY$, and so in this case the map
\begin{equation*}
\cvl : C^{k+2,\alpha} _{-\delta,-\nu} (M;TM) \oplus \YY \rightarrow C^{k,\alpha} _{-\delta,-\nu-2} (M;TM)
\end{equation*}

\noindent is surjective. 

Proving the analogous statement in the AP case requires more work. First observe that one only needs to show that $\cvl : C^{k+2,\alpha} _{\delta,-\nu} \to C^{k,\alpha} _{\delta,-\nu-2}$ is Fredholm whenever $\delta$ is not an indicial root. This is because when $\delta > 0$, integration by parts still implies that $\cvl : C^{k+2,\alpha} _{-\delta,-\nu} \to C^{k,\alpha} _{-\delta-\nu-2}$ is injective, and Fredholmness implies that the cokernel of this map is finite-dimensional. On the other hand, choosing $\epsilon > 0$ so small that there are no indicial roots of $\cvl$ with imaginary part in the interval $[-\delta,-\delta + \epsilon]$, it was shown in \cite{cmp} that $\cvl$ is surjective as a map 
\[
\cvl : H^{k+2} _{-\delta + \epsilon,-\nu+\epsilon} (M;TM) \to H^k _{-\delta + \epsilon,-\nu-2-\epsilon} \supset C^{k,\alpha} _{-\delta,-\nu-2}(M;TM).
\]

\noindent It thus follows that there is some finite-dimensional subset $\YY' \subset H^{k+2} _{-\delta + \epsilon}$ such that the map
\[
\cvl : C^{k+2,\alpha} _{-\delta,-\nu} (M;TM) \oplus \YY' \to C^{k,\alpha} _{-\delta,-\nu-2} (M;TM)
\] 

\noindent is surjective. After the proof of Theorem \ref{holdfredholm} below, we will explain why $\YY' = 0$ in general.

We now prove that the map $\cvl : C^{k+2,\alpha} _{-\delta} \to C^{k,\alpha} _{-\delta}$ is Fredholm on a manifold with only AP ends whenever $\delta$ is not the imaginary part of an indicial root. The parametrix gluing construction described in \cite{cmp} obviates how one extends this result to prove Corollary \ref{trumpetcvlcor}. The proof was communicated to the author by Rafe Mazzeo, and a detailed treatment can be found in \cite{lea1}. Any errors are the author's.

\begin{thm}\label{holdfredholm}

If $(M,g)$ is a Riemannian manifold with a finite number of AP ends, and $\delta \in \RR$ is not an indicial root of $\cvl$, then the map $\cvl : C^{k+2,\alpha} _\delta (M;TM) \to C^{k,\alpha} _\delta (M;TM)$ is Fredholm.

\end{thm}

\proof We show that $\ker (\cvl)$ and $\coker (\cvl)$ are finite dimensional and that $\cvl$ has closed range.
\begin{enumerate}

	\item $\ker (\cvl)$ is finite dimensional: Defining $\epsilon > 0$ as above, we see that $C^{k+2,\alpha} _\delta \subset H^{k+2} _{\delta + \epsilon}$. Since it was shown in \cite{cmp} that $\ker (\cvl |_{H^{k+2} _{\delta + \epsilon}} )$ is finite dimensional, it is immediate that $\ker (\cvl |_{C^{k+2,\alpha} _{\delta}})$ is finite dimensional is as well.
	
	\item $\coker (\cvl)$ is finite dimensional: We argue by contradiction, and suppose there is an infinite set of linearly independent vector fields $V_j$ such that $\|V_j\|_{k,\alpha,\delta} = 1$ and $\mathrm{dist}(V_j,\ran (\cvl|_{C^{k+2,\alpha}_\delta})) \geq \frac12$. Since it was shown in \cite{cmp} that the codimension of $\ran (\cvl : H^{k+2} _{\delta + \epsilon} \to H^k _{\delta + \epsilon})$ is finite, we may assume (taking linear combinations, if necessary) that there exists some $m$ such that $V_j \perp (\ran (\cvl|_{L^2 _{\delta + \epsilon}}))$ for $1 \leq j \leq m$ and $V_j \in (\ran (\cvl|_{L^2 _{\delta + \epsilon}}))$ for $j > m$. Hence for $j > m$, there exists $X_j \in H^{k+2} _{\delta + \epsilon}$ for which $\cvl X_j = V_j$ and $\|X_j\|_{L^2 _{\delta +\epsilon}} \leq C$. Now on each period $[x,x+T]$, the $L^2$-norm and $C^{k+2,\alpha}$-norm are compatible, and so $\|X_j\|_{k+2,\alpha,\delta+\epsilon} \leq C'$.
	Now by assumption, $X_j \notin C^{k+2,\alpha} _\delta$ for $j > m$. We then define a slide-back sequence by setting
	\[
	Y_j(x,y) = e^{-N_j T (\delta + \epsilon)}X_j(x-N_j T,y)
	\]
	
	\noindent where we have chosen $N_j$ such that $|Y_j(x_j,y_j)| \geq \frac12$ and $N_j T \leq x_j \leq (N_j + 1)T$, and also set
	\[
	W_j(x,y) = e^{-N_j T(\delta + \epsilon)}V_j (x - N_j T,y).
	\]
	
	\noindent If we let $\cvl^{(j)}$ denote the operator $\cvl$ with its coefficients translated by $N_j T$ in the $x$-direction, we see that $\cvl^{(j)} Y_j = W_j$, and taking $j \to \infty$ we find that locally $Y_j \to Y$ in $C^{k+2,\alpha}$, where $Y$ is a nontrivial vector field on the exactly periodic cylinder which satisfies $-\mr\Delta_\mathbb{L} Y = 0$ and $\|Y\|_{C^{k+2,\alpha} _{\delta + \epsilon} (\Sigma \times \RR)} \leq C$. However, since $X_j \notin C^{k+2,\alpha} _\delta$, it follows that $Y \notin C^{k+2,\alpha} _\delta (\Sigma \times \RR)$, but this contradicts the fact that we defined $\epsilon$ to be so small that no indicial roots have imaginary part lying between $\delta$ and $\delta + \epsilon$. This contradiction proves that the cokernel of the map
	\[
	\cvl : C^{k+2,\alpha} _\delta (M;TM) \to C^{k,\alpha} _\delta (M;TM)
	\]
	
	is finite dimensional.
	
	\item $\cvl$ has closed range: We argue by contradiction. Suppose that there is a sequence $X_j \in C^{k+2,\alpha} _\delta$ such that $\cvl X_j = V_j$ and $V_j \to V$ in $C^{k,\alpha} _\delta$ but the sequence $X_j$ does not converge in $C^{k+2,\alpha} _\delta$. Without loss of generality, assume that the sequence $\{X_j\}$ lies in some fixed complement $\mathcal{H}$ of $\ker (\cvl)$. If $\|X_j\|_{k+2,\alpha,\delta}$ stayed bounded, one could apply the Arzel\`a-Ascoli Theorem to extract a convergent subsequence, and so, passing to a subsequence if necessary, we must have $\|X_j\|_{k+2,\alpha,\delta} = A_j \to \infty$. This and local Schauder estimates for $\cvl$ imply that we may find points $(x_j,y_j)$ of $M$ for which $|X_j (x_j,y_j)| \geq \frac12 A_j e^{\delta x_j}$. We first assume that $(x_j,y_j)$ stays bounded. We then define new vector fields $Y_j = X_j / A_j$ so that $\|Y_j\|_{k+2,\alpha,\delta} \leq 1$ and $\sup |Y_j| \geq c \geq 0$ in a fixed compact set. Now clearly $\cvl Y_j = V_j / A_j \to 0$ and therefore the Arzel\`a-Ascoli Theorem again implies that there is some $Y \neq 0$ such that $Y_j \to Y$ in any compact set (with respect to the induced $C^{k+2,\alpha} _\delta$-norm) while $\cvl Y = 0$ and hence $Y \notin \mathcal{H}$, a contradiction.
	
	Now assume that $x_j \to \infty$. We define auxiliary vector fields $Y_j$ by both scaling and translating, where we set
	\[
	Y_j (x,y) = A_j^{-1}e^{-N_j T \delta} X_j (x-N_jT,y), 
	\]
	
	\noindent where we recall that $T$ is the period of the limiting metric and we assume that $N_j T \leq x \leq (N_j + 1)T$. If we similarly define
	\[
	W_j (x,y) = A_j^{-1}e^{-N_j T \delta} V_j (x-N_jT,y),
	\]
	
	\noindent we see that $\cvl Y_j = W_j \to 0$ in $C^{k,\alpha} _\delta$, so here we again see that $Y_j \to Y \neq 0$ with $\cvl Y = 0$ in compact sets. Since we again have a contradiction, we conclude that $\cvl : C^{k+2,\alpha} _\delta \to C^{k,\alpha} _\delta$ has closed range, proving the theorem. \qed
	
\end{enumerate}

\vspace{1pc}


A very similar argument to that above can be used to prove that, whenever $\delta$ is not an indicial root of the operator $\cvl$, we have that $\cvl : C^{k+2,\alpha} _\delta \to C^{k,\alpha} _\delta$ is surjective. In particular, we know that for any $V \in C^{k,\alpha} _\delta \subset H^k _{\delta + \epsilon}$, there is some $X \in H^{k+2} _{\delta +\epsilon}$ such that $\cvl X = V$. Since we are again assuming that $\epsilon > 0$ is so small that there are no indicial roots between $\delta$ and $\delta + \epsilon$, it is straightforward to show that we may choose the same $X$ for any such choice of $\epsilon$. Now we fix some $\epsilon$ and again define
\[
Y_j (x,y) = e^{-N_j T (\delta + \epsilon)} X(x - N_j T,y).
\]

\noindent One argues exactly as in the proof of Theorem \ref{holdfredholm} to show that if $X \notin C^{k+2,\alpha} _\delta$, then there exists some $Y$ on the exactly periodic cylinder which satisfies $- \mr\Delta_\mathbb{L} Y = 0$ whose growth rate on one end is faster than $e^{\delta x}$ but slower than $e^{(\delta + \epsilon)x}$, a contradiction. 

Now suppose $\delta > 0$. To show that $\YY' = 0$, we observe that  $X \in H^{k+2} _{-\delta - \epsilon} \subset H^{k+2} _{-\delta + \epsilon}$. $X$ is independent of $\epsilon$, and so we argue just as in the previous paragraph to show that $X' \in C^{k+2,\alpha} _{-\delta}$, and we may thus take $\YY' = 0$.


\bibliographystyle{abbrv}
\bibliography{trumpet_data}

\end{document}